\documentclass[prl,twocolumn,showpacs,preprintnumbers,amsmath,amssymb]{revtex4}
\usepackage{colordvi}
\usepackage{amssymb}
\usepackage{graphics}
\usepackage{pstricks}
\usepackage{fancybox}
\usepackage{rotating}
\usepackage{epsfig}

\newcommand{\gtap}{\;{\raise.3ex\hbox{$>$\kern-.75em\lower1ex\hbox{$\sim$}}}\;}
\newcommand{\ltap}{\;{\raise.3ex\hbox{$<$\kern-.75em\lower1ex\hbox{$\sim$}}}\;}

\begin{document}

\preprint{CP3-07-08, DSF-6/2007}

\title{QCD corrections to $J/\psi$ and $\Upsilon$ production at hadron colliders} 

\author{J.~Campbell$^1$, F.~Maltoni$^2$, F.~Tramontano$^3$}
\affiliation{$^1$ Department of Physics and Astronomy, University of Glasgow, 
Glasgow G12 8QQ, United Kingdom \\
$^2$ Center for Particle Physics and Phenomenology (CP3),
Universit\'{e} Catholique de Louvain, B-1348 Louvain-la-Neuve, Belgium\\
$^3$ Universit\`a di Napoli Federico II, Dipartimento di Scienze Fisiche, and INFN, Sezione di Napoli, I-80126 Napoli, Italy}

\begin{abstract}
We calculate the cross section for hadroproduction of a pair of heavy
quarks in a $^3S_1$ color-singlet state at next-to-leading order in
QCD. This corresponds to the leading contribution in the NRQCD
expansion for $J/\psi$ and $\Upsilon$ production. The higher-order
corrections have a large impact on the $p_T$ distributions, enhancing
the production at high $p_T$ both at the Tevatron and at the LHC. 
The total decay rate of a $^3S_1$ into hadrons at NLO is 
also computed, confirming for the first time the result obtained 
by Mackenzie and Lepage in 1981.
\end{abstract}


\maketitle

%introduction
{\bf 1.} Charmonium and bottomonium states are certainly among
the most interesting systems to test our understanding 
of the strong interactions, both at the perturbative and 
non-perturbative level. More than
thirty years after the discovery of the first charm-anticharm
resonance, the $J/\psi$, the study of their properties, including
production and decay mechanisms, is still the subject of considerable
interest~\cite{Brambilla:2004wf}.  

From the theoretical point of view, a rigorous framework, based on the
use of non-relativistic QCD (NRQCD)~\cite{Bodwin:1994jh}, has been
introduced that allows consistent theoretical predictions to be made
and to be systematically improved.  However, despite theoretical
developments and successes, not all the predictions of the NRQCD
factorization approach have been firmly established. Recent
measurements in $e^+e^-$ collisions have shown that production rates
for single and double charmonium production are in general much larger
than those predicted by leading order
calculations~\cite{Bodwin:2005ec}. Measurements at the Tevatron in
proton-antiproton collisions are not fully compatible with those obtained
at HERA in electron-proton collisions~\cite{Kramer:2001hh} and in
fixed-target experiments~\cite{maltoni:2006yp}, suggesting the
possibility that charmonium might be too light for the NRQCD factorization and/or scaling rules to work.
In this context, a real challenge is offered by measurements of the
$J/\psi$\ polarization at the Tevatron. NRQCD predicts a sizeable
transverse polarization for $J/\psi$ 's at high-$p_T$, in contrast
with the latest data that now clearly indicate that $J/\psi$'s are
not transversely polarized~\cite{cdf-note}.

In view of such a puzzling scenario, it is worth to re-examine in
detail the theoretical predictions and try to systematically improve
on them.  In this Letter we report on the calculation of the
next-to-leading order correction to the hadroproduction of a pair of
heavy quarks in a color-singlet $^3S_1$ state. This is the leading term
in the NRQCD expansion for a $J/\psi$ or a $\Upsilon$ and it is
equivalent to the ``color-singlet model'' approximation. This model assumes
that the non-perturbative dynamics leaves unchanged the quantum
numbers of the perturbative quark-antiquark pair, which are the same as
those of the physical bound state.  It is curious to note that
the analogous corrections to the hadronic decay width of a color-singlet $^3S_1$ state, which involve
exactly the same diagrams, have been available for a long
time~\cite{Mackenzie:1981sf} and play an important role in the
extraction of $\alpha_S$ from $\Upsilon$ decays~\cite{Brambilla:2007cz}. 
Our calculation for the inclusive decay rate at NLO 
provides the first independent confirmation of those results.
\begin{figure}[b]
\begin{center}
\epsfxsize=8cm 
\epsfbox{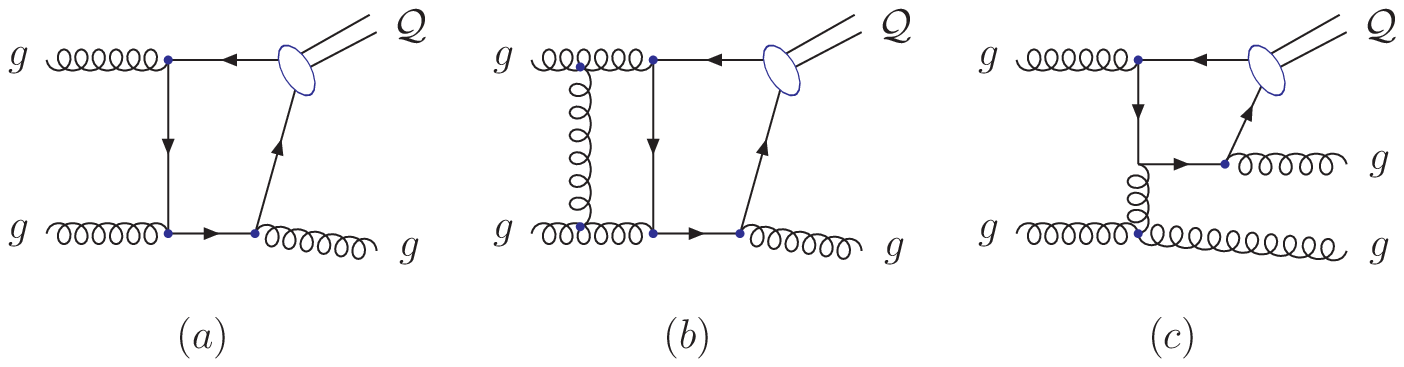}
\vspace*{-.9cm}
\end{center}
\caption{Representative Feynman diagrams for $^3S_1^{[1]}$ hadroproduction
at LO (a),  virtual (b), and real (c) contributions at NLO. 
Amplitudes with a light quark line (not shown) also 
contribute to the real corrections.}
\label{fig:production}
\end{figure}

{\bf 2.} %production
According to the NRQCD factorization 
approach~\cite{Bodwin:1994jh}, 
the inclusive cross section for direct production 
for a $J/\psi$, $\Upsilon$ and
their radial excitations, (hereafter denoted $\cal{Q}$) in hadron-hadron 
collisions can be written as:
\begin{eqnarray}
\sigma( p p  \to {\cal Q}  + X) &=&  
\sum_{i,j,n} \int dx_1 dx_2 f_{i/p} f_{j/p}\,   \\
&\times& \hat  \sigma[ij \to (Q\bar Q)_n + x ] 
\langle 0| {\cal O}^{\cal Q}_n|0 \rangle\nonumber \,, 
\end{eqnarray}
where $p$ is either a proton or an antiproton.
The short-distance coefficients $\hat \sigma$ are calculable in
perturbative QCD and describe the production of a heavy quark--antiquark pair
$Q\bar Q$ state with quantum numbers $n$. Conversely, the $\langle 0 | {\cal
O}^{\cal{Q}}_n |0 \rangle$ are the non-perturbative matrix elements
that describe the subsequent hadronization of the $Q\bar Q$ pair into
the physical $\cal{Q}$ state.  These matrix elements can be expanded
in powers of the relative velocity of the heavy quarks in the bound
state, $v^2$. Since $v^2 \simeq 0.3 (0.1)$ for charmonium (bottomonium)
only a few terms need to be included in the sum over
$n$ to reach a given accuracy.
Compared to the leading contribution, the $^3S_1^{[1]}(Q\bar
Q)$, the color-octet terms, ($^1S_0^{[8]},^3S_1^{[8]},^3P_J^{[8]}$),
are all suppressed by $v^4$.  However, the corresponding short
distance coefficients might compensate this suppression, either by
having leading order contributions starting at $\alpha_S^2$ compared
to the $\alpha_S^3$ of the singlet and/or by kinematic enhancements due
to a different scaling of the differential cross section with the
$p_T$.  In particular it is easy to verify that the partonic
differential cross sections at LO behave as $1/p_T^6$ for CP-even
color-octet states and $1/p_T^4$ for the $^3S_1^{[8]}$, compared to
the $1/p_T^8$ scaling of the color-singlet.  In the case of charmonium
production at the Tevatron, the $1/p_T^4$ and $1/p_T^6$ behaviors are
clearly seen in the data~\cite{Abe:1997jz} and the introduction of the contribution from
the octets allows an explanation of the observed $p_T$ spectrum~\cite{Beneke:1996yw}. 
On the other hand, at NLO $^3S_1^{[1]}$ production has both $1/p_T^6$ and $1/p_T^4$
components, the latter coming from the associated production with a
heavy-quark pair, that can affect the interpretation of these
behaviors in terms of color-octet matrix elements~\cite{Petrelli:1999rh}.

{\bf 3.} Analogously to orthopositronium in QED, for a $^3S_1$ state
the first non-zero contribution in QCD comes from the coupling to
three gauge vectors, so that in hadron collisions the quarkonium is produced
in association with a gluon, Fig.~\ref{fig:production}(a).  Since we
want to retain spin-correlations and work with helicity amplitudes, we
find it convenient to include the decay of the quarkonium state in to a
pair of leptons, $\ell \bar \ell$.  In so doing the projection of the
quantum numbers (color and spin) of the heavy-quark pair to a
$^3S_1^{[1]}$ state is automatic.  The color-ordered helicity
amplitudes for the short distance process $ggg \to ^3S_1^{[1]} \to \ell \bar
\ell$ are
\begin{eqnarray}
&&A(1^+,2^+,3^+;5_\ell^+,6_{\bar \ell}^-)= 0 \,,\nonumber \\
&&A(1^+,2^+,3^-;5_\ell^+,6_{\bar \ell}^-)= \nonumber \\
&&{\cal N} \frac{\langle 3 5 \rangle^2 [12]^2 [56]}
{(s_{12}+s_{13}) (s_{13}+s_{23}) (s_{12}+s_{23}) }\,,
\end{eqnarray}
where ${\cal N}= 8 \sqrt{2} g_s^3 e^2 m_Q$. 

Next-to-leading order corrections include virtual corrections to the
$2 \to 2$ process, Fig.~\ref{fig:production}(b), the real
processes, such as the purely gluonic contribution $gg\to {\cal Q}
gg$, Fig.~\ref{fig:production}(c). In addition, channels with a
quark-antiquark pair: $ gq \to {\cal Q} g q $ and their crossings
also contribute.

The amplitudes $gggg,ggq\bar q \to ^3S_1^{[1]} \to \ell \bar \ell$
which enter the real corrections, have been calculated analytically
and checked numerically against those obtained with
MadGraph~\cite{Stelzer:1994ta}. We note that by squaring the full
amplitudes involving four gluons and summing over color we find a
non-trivial simplification, with only one kinematical combination.
The overall color coefficient is given by $N_c C_2(F)$, with $C_2(F)$
being the second Casimir of $SU(3)$ and corresponding to the color
factor of the Born squared amplitude, i.e., $C_2(F)=(d^{abc}/\sqrt{N_c})^2$.

Virtual corrections to the leading order helicity amplitudes have been
calculated within the four dimensional helicity scheme. Dimensional
regularization has been adopted for isolating the ultraviolet,
infrared and collinear singularities. The heavy-quark mass and wave
function have been renormalized on-shell.  $\overline{ MS}$
renormalization has been employed for the gluon wave function and
$\alpha_S$ renormalization.  The color projection onto a singlet
state leads to remarkable simplifications at NLO.
We split diagrams containing the four-gluon vertex into a sum of color ordered terms,
where each term is written as a product of color matrices in the fundamental
representation, $\lambda^a$. In this way each
virtual diagram contributes to all of the six gluon permutations
each multiplying ${\rm Tr}(\lambda^{a} \lambda^{b} \lambda^{c})$.
In fact, out of the above six permutations,
only one independent combination is left which is the sum of three
permutations with the same order of the color generators.  The full
amplitudes are proportional to $d^{abc}/\sqrt{N_c}$, exactly as for the
Born amplitude. Thus all the contributions from individual diagrams for each
helicity configuration can be summed together, resulting in a sizeable
simplification of analytic expressions for the final results. We also
find that the infrared divergences are proportional to $N_c$, while
the subleading terms in $N_c$ are finite, 
in agreement with the simplifications mentioned above
for the real corrections.  As a further check of our calculation, we
verified numerically the identity of the two color ordered
subamplitudes and their gauge invariance under different choices for
the polarization vectors of the three gluons.

To extract the singularities of the real part of the NLO corrections
$\sigma^{\mathrm{real}}$, the dipole subtraction formalism
\cite{Catani:1996vz} has been adopted, following the implementation
of MCFM~\cite{Campbell:1999ah}.  The algorithm is based on
mapping the singularities of the real cross section
$\sigma^{\mathrm{real}}$ onto an auxiliary cross section
$\sigma^{\mathrm{sub}}$ which can be built out of universal building
blocks, the so-called dipoles, and it is simple enough that the
singular regions in phase space can be integrated out
analytically. The remaining piece,
$\sigma^{\mathrm{real}}-\sigma^{\mathrm{sub}}$ is finite and can be
safely integrated with standard Monte Carlo techniques.  The
divergences in the auxiliary cross section $\sigma^{\mathrm{sub}}$ are
in part canceled by the the soft and collinear singularities of the
virtual corrections while the rest are absorbed in the renormalization
of the parton densities. Note that no dipoles for the heavy quarks are
needed since there are no infrared divergences associated with a
collinear heavy quark-antiquark pair in a color-singlet state. Indeed
we find that the structure of the infrared divergences is the same as
that of Higgs hadroproduction in association with a
jet~\cite{Glosser:2002gm}, as it should be.
\begin{figure}[t!]
\begin{center}
\begin{sideways}
\epsfxsize=5cm 
\epsfbox{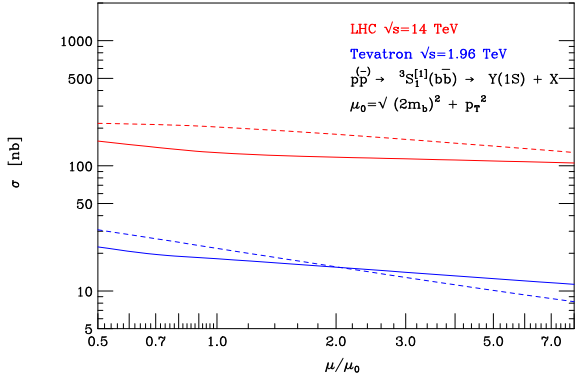}
\end{sideways}
 \vspace*{0cm}
\end{center}
\caption{Variation of the cross section for direct $\Upsilon$ 
production at the Tevatron (lower curves) and the LHC (upper curves) 
at LO (dashes) and NLO (solid).  Renormalization and factorization scales are set equal, $\mu_F=\mu_R$, and $\mu_0=\sqrt{(2 m_b)^2+p_T^2}$.}
\label{fig:b-scale}
\end{figure}
%
% results
%
\begin{figure}[t!]
\begin{center}
\begin{sideways}
\epsfxsize=5cm 
\epsfbox{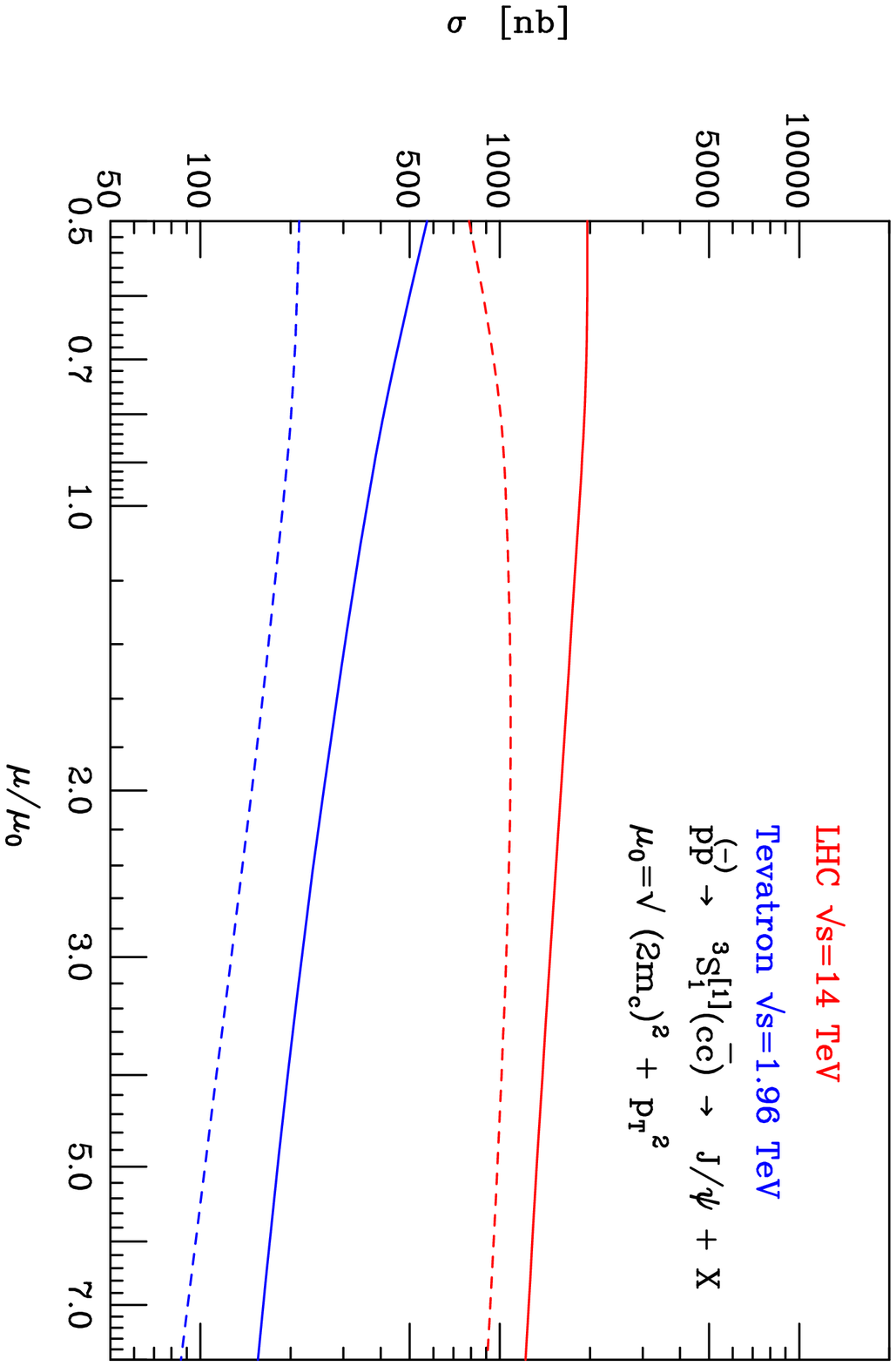}
\end{sideways}
 \vspace*{0cm}
\end{center}
\caption{Variation of the cross section for direct $J/\psi$ 
production at the Tevatron (lower curves) and the LHC (upper curves) 
at LO (dashes) and NLO (solid). Renormalization and factorization scales are set equal $\mu_F=\mu_R$, and $\mu_0=\sqrt{(2 m_c)^2+p_T^2}$.}
\label{fig:c-scale}
\end{figure}

\begin{figure}[t!]
\begin{center}
\begin{sideways}
\includegraphics[width=5cm]{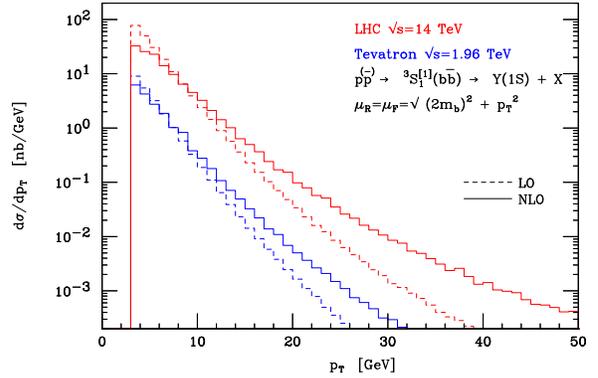}
\end{sideways}
 \vspace*{0cm}
\end{center}
\caption{Differential cross sections for direct $\Upsilon$ 
production via a $^3S_1^{[1]}$ intermediate state, at the Tevatron (lower histograms) and LHC (upper histograms), at LO (dashes) and NLO (solid). $p_T^{\Upsilon}>3$ GeV and $|y^{\Upsilon}|<3$. Details on the input 
parameters are given in the text.}
\label{fig:b-pt}
\end{figure}
\vspace*{0.5em}
\noindent
{\bf 4.} We now turn to the presentation of the main features of the NLO
results. In our numerical studies, the  
CTEQ6L1 (CTEQ6M) parton distribution functions and the corresponding fitted value 
for $\alpha_S(M_Z)=0.130 (0.118)$ are used for the LO (NLO) predictions.
The heavy-quark mass and the color-singlet non-perturbative matrix element are 
$\langle {\cal O}^{\Upsilon}_1(^3S_1)\rangle = 9.28$ GeV$^3$ and 
$m_b=4.75$ GeV for the $\Upsilon$, and 
$\langle {\cal O}^{J/\psi}_1(^3S_1)\rangle = 1.16$ GeV$^3$ and $m_c=1.5$ GeV for
the $J/\psi$.

The inclusion of the NLO corrections leads to important modifications
of the total cross sections and the distributions. 
First we note that, even though the calculation of the total cross
section at NLO is infrared-safe, there are some regions of the phase
space, notably at low $p_T$ and at large rapidity of the quarkonium
state, where the perturbative expansion is not under proper control
and a fixed-order approximation fails to provide a reliable estimate. 
This is the same behavior found in the corresponding regions of the phase
space in the calculation for $\gamma p \to$
$^3S_1^{[1]} +X $ at NLO~\cite{Kramer:1995nb}. A careful treatment of
the effects of soft radiation as outlined, {\it e.g.}, in Ref.~\cite{Berger:2004cc}
is beyond the scope of this Letter and is left for future work. For the
sake of illustration, we restrict our results to the domain $p_T^{\cal
Q} >3$ GeV and $|y^{\cal Q}|<3$ where the cross section is always
positive definite. We then study how the calculation behaves by
looking at the dependence of the cross section on variations of the
renormalization and factorization scales between $\mu_0/2
<\mu_F=\mu_R<2 \mu_0$, with $\mu_0=\sqrt{(2 m_{\cal Q})^2+p_T^2}$. The results are shown in
Figs.~\ref{fig:b-scale} and ~\ref{fig:c-scale}.  We find that the NLO
prediction for $\Upsilon$ production is very well behaved both at
Tevatron and the LHC and the scale dependence is (slightly) improved.
For $J/\psi$ the scale dependence at NLO is not improved in this region of the phase space.

\begin{figure}[t!]
\begin{center}
\begin{sideways}
\includegraphics[width=5cm]{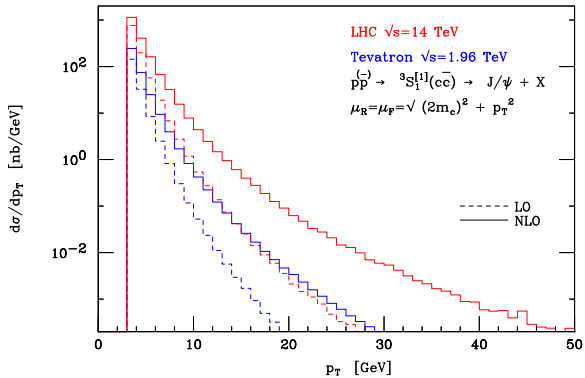}
\end{sideways}
 \vspace*{0cm}
\end{center}
\caption{Differential cross sections for direct $J/\psi$ 
production via a $^3S_1^{[1]}$ intermediate state, at the Tevatron (lower histograms) and LHC (upper histograms), at LO (dashes) and NLO (solid). $p_T^{J/\psi}>3$ GeV and $|y^{J/\psi}|<3$. Details on the input 
parameters are given in the text.}
\label{fig:c-pt}
\end{figure}

In Figs.~\ref{fig:b-pt} and ~\ref{fig:c-pt}  we plot the inclusive 
$p_T$ distributions of the quarkonium states, with $p_T^{\cal Q}>3$ GeV and $|y^{\cal Q}|<3$.
The factorization and renormalization scales are chosen as $\mu_F=\mu_R=\sqrt{(2 m_{\cal Q})^2+p_T^2}$.
As expected, the NLO curves have a much milder
drop than the LO ones, due to different $p_T$ scaling of the real
contributions.  This feature leads to much larger cross sections 
with respect to those predicted at LO, reaching  
one order of magnitude increase for $p_T^{\cal Q} \gtrsim 4 (2 m_Q)$.

As an important check and application of our results we also
calculated the NLO corrections to the decay width of a $^3S_1^{[1]}$
into hadrons. Differences with the production cross sections are minor
and simply amount to the analytic continuation of the loop integral
to the physical region relevant to the decay process and the
integration over 3-body and 4-body phase spaces for the virtual and
real contributions, respectively. Our result is
\begin{widetext}
\begin{eqnarray}
&&\Gamma^{\mbox{NLO}}(\Upsilon\to LH)
=\Gamma^{\mbox{LO}}\,\left[ 1+\frac{\alpha_S(\mu)}{\pi} % \right.  \nonumber\\ 
%&&\times  \left. 
\left(
       - 9.471 ~ C_F 
       + 4.106 ~ C_A 
%\right.\right.\\&&\left.\left. ~~    
       - 1.150 ~ n_f + 
       \frac{3}{2} \beta_0  \log \frac{\mu}{m_b}
 \right)\right]\,,
 \label{eq:decay}
\end{eqnarray}
\end{widetext}
with $\beta_0=\frac{11}{3} C_A - \frac23 n_f$ and $C_F=4/3,C_A=3$. 
The numerical coefficients, obtained by a Monte Carlo integration
method, have an uncertainty better than one per mil, and are in good
agreement with those of Ref.~\cite{Mackenzie:1981sf}.
Eq.~(\ref{eq:decay}) provides the first independent check of the
results of that seminal work.

% conclusions

\vspace*{0.5em}
\noindent
{\bf 5.} To summarize, we have presented the NLO cross section for the
production of a heavy quark pair in a $^3S_1^{[1]}$ state at hadron
colliders. This is the leading contribution in the relative velocity
expansion to the production of $J/\psi$, $\Upsilon$ and their radial
excitations. The NLO corrections are large and modify the expectations
for the total cross sections and the distribution in the transverse
momentum. Further analysis, including combination with the results of
Ref.~\cite{Artoisenet1} for $pp \to {\cal Q} Q\bar Q$ 
and Ref.~\cite{Petrelli:1997ge}, to compare
with the available data from the Tevatron and fixed-target experiments
is the subject of a forthcoming paper.

\vspace*{1.5em}
\noindent
{\bf Acknowledgments} We would like to thank Michael Kr\"amer and 
Andrea Petrelli for their collaboration during the initial stages 
of this project, and Carlo Oleari for useful discussions. F.M. is 
partially supported by the Belgian Federal Science Policy (IAP 6/11).

\bibliography{database}

\begin{thebibliography}{19}
\expandafter\ifx\csname natexlab\endcsname\relax\def\natexlab#1{#1}\fi
\expandafter\ifx\csname bibnamefont\endcsname\relax
  \def\bibnamefont#1{#1}\fi
\expandafter\ifx\csname bibfnamefont\endcsname\relax
  \def\bibfnamefont#1{#1}\fi
\expandafter\ifx\csname citenamefont\endcsname\relax
  \def\citenamefont#1{#1}\fi
\expandafter\ifx\csname url\endcsname\relax
  \def\url#1{\texttt{#1}}\fi
\expandafter\ifx\csname urlprefix\endcsname\relax\def\urlprefix{URL }\fi
\providecommand{\bibinfo}[2]{#2}
\providecommand{\eprint}[2][]{\url{#2}}

\bibitem[{\citenamefont{Brambilla et~al.}(2004)}]{Brambilla:2004wf}
\bibinfo{author}{\bibfnamefont{N.}~\bibnamefont{Brambilla}}
  \bibnamefont{et~al.} (\bibinfo{year}{2004}), \eprint{hep-ph/0412158}.

\bibitem[{\citenamefont{Bodwin et~al.}(1995)\citenamefont{Bodwin, Braaten, and
  Lepage}}]{Bodwin:1994jh}
\bibinfo{author}{\bibfnamefont{G.~T.} \bibnamefont{Bodwin}},
  \bibinfo{author}{\bibfnamefont{E.}~\bibnamefont{Braaten}}, \bibnamefont{and}
  \bibinfo{author}{\bibfnamefont{G.~P.} \bibnamefont{Lepage}},
  \bibinfo{journal}{Phys. Rev.} \textbf{\bibinfo{volume}{D51}},
  \bibinfo{pages}{1125} (\bibinfo{year}{1995}), \eprint{hep-ph/9407339}.

\bibitem[{\citenamefont{Bodwin}(2006)}]{Bodwin:2005ec}
\bibinfo{author}{\bibfnamefont{G.~T.} \bibnamefont{Bodwin}},
  \bibinfo{journal}{Int. J. Mod. Phys.} \textbf{\bibinfo{volume}{A21}},
  \bibinfo{pages}{785} (\bibinfo{year}{2006}), \eprint{hep-ph/0509203}.

\bibitem[{\citenamefont{Kramer}(2001)}]{Kramer:2001hh}
\bibinfo{author}{\bibfnamefont{M.}~\bibnamefont{Kramer}},
  \bibinfo{journal}{Prog. Part. Nucl. Phys.} \textbf{\bibinfo{volume}{47}},
  \bibinfo{pages}{141} (\bibinfo{year}{2001}), \eprint{hep-ph/0106120}.

\bibitem[{\citenamefont{Maltoni et~al.}(2006)}]{maltoni:2006yp}
\bibinfo{author}{\bibfnamefont{F.}~\bibnamefont{Maltoni}} \bibnamefont{et~al.},
  \bibinfo{journal}{Phys. Lett.} \textbf{\bibinfo{volume}{B638}},
  \bibinfo{pages}{202} (\bibinfo{year}{2006}), \eprint{hep-ph/0601203}.

\bibitem[{\citenamefont{CDF}()}]{cdf-note}
\bibinfo{author}{\bibnamefont{CDF}}, \bibinfo{note}{collaboration, Note 8212,
  06-06-22}.

\bibitem[{\citenamefont{Mackenzie and Lepage}(1981)}]{Mackenzie:1981sf}
\bibinfo{author}{\bibfnamefont{P.~B.} \bibnamefont{Mackenzie}}
  \bibnamefont{and} \bibinfo{author}{\bibfnamefont{G.~P.}
  \bibnamefont{Lepage}}, \bibinfo{journal}{Phys. Rev. Lett.}
  \textbf{\bibinfo{volume}{47}}, \bibinfo{pages}{1244} (\bibinfo{year}{1981}).

\bibitem[{\citenamefont{Brambilla et~al.}(2007)\citenamefont{Brambilla, Tormo,
  Soto, and Vairo}}]{Brambilla:2007cz}
\bibinfo{author}{\bibfnamefont{N.}~\bibnamefont{Brambilla}},
  \bibinfo{author}{\bibfnamefont{X.~G.~i.} \bibnamefont{Tormo}},
  \bibinfo{author}{\bibfnamefont{J.}~\bibnamefont{Soto}}, \bibnamefont{and}
  \bibinfo{author}{\bibfnamefont{A.}~\bibnamefont{Vairo}}
  (\bibinfo{year}{2007}), \eprint{hep-ph/0702079}.

\bibitem[{\citenamefont{Abe et~al.}(1997)}]{Abe:1997jz}
\bibinfo{author}{\bibfnamefont{F.}~\bibnamefont{Abe}} \bibnamefont{et~al.}
  (\bibinfo{collaboration}{CDF}), \bibinfo{journal}{Phys. Rev. Lett.}
  \textbf{\bibinfo{volume}{79}}, \bibinfo{pages}{572} (\bibinfo{year}{1997}).

\bibitem[{\citenamefont{Beneke and Kramer}(1997)}]{Beneke:1996yw}
\bibinfo{author}{\bibfnamefont{M.}~\bibnamefont{Beneke}} \bibnamefont{and}
  \bibinfo{author}{\bibfnamefont{M.}~\bibnamefont{Kramer}},
  \bibinfo{journal}{Phys. Rev.} \textbf{\bibinfo{volume}{D55}},
  \bibinfo{pages}{5269} (\bibinfo{year}{1997}), \eprint{hep-ph/9611218}.

\bibitem[{\citenamefont{Petrelli}(2000)}]{Petrelli:1999rh}
\bibinfo{author}{\bibfnamefont{A.}~\bibnamefont{Petrelli}},
  \bibinfo{journal}{Nucl. Phys. Proc. Suppl.} \textbf{\bibinfo{volume}{86}},
  \bibinfo{pages}{533} (\bibinfo{year}{2000}), \eprint{hep-ph/9910274}.

\bibitem[{\citenamefont{Stelzer and Long}(1994)}]{Stelzer:1994ta}
\bibinfo{author}{\bibfnamefont{T.}~\bibnamefont{Stelzer}} \bibnamefont{and}
  \bibinfo{author}{\bibfnamefont{W.~F.} \bibnamefont{Long}},
  \bibinfo{journal}{Comput. Phys. Commun.} \textbf{\bibinfo{volume}{81}},
  \bibinfo{pages}{357} (\bibinfo{year}{1994}), \eprint{hep-ph/9401258}.

\bibitem[{\citenamefont{Catani and Seymour}(1997)}]{Catani:1996vz}
\bibinfo{author}{\bibfnamefont{S.}~\bibnamefont{Catani}} \bibnamefont{and}
  \bibinfo{author}{\bibfnamefont{M.~H.} \bibnamefont{Seymour}},
  \bibinfo{journal}{Nucl. Phys.} \textbf{\bibinfo{volume}{B485}},
  \bibinfo{pages}{291} (\bibinfo{year}{1997}), \eprint{hep-ph/9605323}.

\bibitem[{\citenamefont{Campbell and Ellis}(1999)}]{Campbell:1999ah}
\bibinfo{author}{\bibfnamefont{J.~M.} \bibnamefont{Campbell}} \bibnamefont{and}
  \bibinfo{author}{\bibfnamefont{R.~K.} \bibnamefont{Ellis}},
  \bibinfo{journal}{Phys. Rev.} \textbf{\bibinfo{volume}{D60}},
  \bibinfo{pages}{113006} (\bibinfo{year}{1999}), \eprint{hep-ph/9905386}.

\bibitem[{\citenamefont{Glosser and Schmidt}(2002)}]{Glosser:2002gm}
\bibinfo{author}{\bibfnamefont{C.~J.} \bibnamefont{Glosser}} \bibnamefont{and}
  \bibinfo{author}{\bibfnamefont{C.~R.} \bibnamefont{Schmidt}},
  \bibinfo{journal}{JHEP} \textbf{\bibinfo{volume}{12}}, \bibinfo{pages}{016}
  (\bibinfo{year}{2002}), \eprint{hep-ph/0209248}.

\bibitem[{\citenamefont{Kramer}(1996)}]{Kramer:1995nb}
\bibinfo{author}{\bibfnamefont{M.}~\bibnamefont{Kramer}},
  \bibinfo{journal}{Nucl. Phys.} \textbf{\bibinfo{volume}{B459}},
  \bibinfo{pages}{3} (\bibinfo{year}{1996}), \eprint{hep-ph/9508409}.

\bibitem[{\citenamefont{Berger et~al.}(2005)\citenamefont{Berger, Qiu, and
  Wang}}]{Berger:2004cc}
\bibinfo{author}{\bibfnamefont{E.~L.} \bibnamefont{Berger}},
  \bibinfo{author}{\bibfnamefont{J.-w.} \bibnamefont{Qiu}}, \bibnamefont{and}
  \bibinfo{author}{\bibfnamefont{Y.-l.} \bibnamefont{Wang}},
  \bibinfo{journal}{Phys. Rev.} \textbf{\bibinfo{volume}{D71}},
  \bibinfo{pages}{034007} (\bibinfo{year}{2005}), \eprint{hep-ph/0404158}.

\bibitem[{\citenamefont{Artoisenet et~al.}()\citenamefont{Artoisenet, Lansberg,
  and Maltoni}}]{Artoisenet1}
\bibinfo{author}{\bibfnamefont{P.}~\bibnamefont{Artoisenet}},
  \bibinfo{author}{\bibfnamefont{J.-P.} \bibnamefont{Lansberg}},
  \bibnamefont{and} \bibinfo{author}{\bibfnamefont{F.}~\bibnamefont{Maltoni}},
  \eprint{hep-ph/0703129}.

\bibitem[{\citenamefont{Petrelli et~al.}(1998)\citenamefont{Petrelli, Cacciari,
  Greco, Maltoni, and Mangano}}]{Petrelli:1997ge}
\bibinfo{author}{\bibfnamefont{A.}~\bibnamefont{Petrelli}},
  \bibinfo{author}{\bibfnamefont{M.}~\bibnamefont{Cacciari}},
  \bibinfo{author}{\bibfnamefont{M.}~\bibnamefont{Greco}},
  \bibinfo{author}{\bibfnamefont{F.}~\bibnamefont{Maltoni}}, \bibnamefont{and}
  \bibinfo{author}{\bibfnamefont{M.~L.} \bibnamefont{Mangano}},
  \bibinfo{journal}{Nucl. Phys.} \textbf{\bibinfo{volume}{B514}},
  \bibinfo{pages}{245} (\bibinfo{year}{1998}), \eprint{hep-ph/9707223}.

\end{thebibliography}
\end{document}